\documentclass[aps]{revtex4}

\usepackage[dvipdfm]{graphicx}
\usepackage{amsmath, amssymb}

\usepackage{color}

\usepackage{ulem}

\begin{document}

\title{Optimal phase response curves for stochastic synchronization of limit-cycle oscillators by common Poisson noise}

\author{Shigefumi Hata$^{1,2}$, Kensuke Arai$^{3}$, Roberto F.~Gal\'an$^{4}$, and Hiroya Nakao$^{2,5}$}

\affiliation{$^{1}$Department of Physics, Kyoto University, Kyoto 606-8502, Japan}
\affiliation{$^{2}$CREST, JST, Kyoto 606-8502, Japan}
\affiliation{$^{3}$Brain Science Institute, RIKEN, Wako 351-0198, Japan}
\affiliation{$^{4}$Department of Neurosciences, Case Western Reserve University, Cleveland, Ohio 44106, USA}
\affiliation{$^{5}$Department of Mechanical and Environmental Informatics, Tokyo Institute of Technology, Tokyo 152-8550, Japan}

\begin{abstract}
We consider optimization of phase response curves for stochastic synchronization of non-interacting limit-cycle oscillators by common Poisson impulsive signals.  The optimal functional shape for sufficiently weak signals is sinusoidal, but can differ for stronger signals.  By solving the Euler-Lagrange equation associated with the minimization of the Lyapunov exponent characterizing synchronization efficiency, the optimal phase response curve is obtained.  We show that the optimal shape mutates from a sinusoid to a sawtooth as the constraint on its squared amplitude is varied.
\end{abstract}

\date{\today}

\maketitle

%%%%%%%%%%%%%%%%%%%%%%%%%%%%%%%%%%%%%%%%%%%%%%%%%%%%%%%%%%%%%%%%%%

\section{Introduction}

Synchronization of non-interacting rhythmic elements by common random driving signals~\cite{Mainen,Uchida,Ranta,Teramae,Goldobin,Goldobin2,Galan1,Nagai,Yoshida,Nakao,Arai,Hata,Toral,Zhou,Pikovsky}, termed {\it stochastic} or {\it noise-induced synchrony}, may explain synchronous behavior of various systems ranging from lasers~\cite{Uchida} and electronic circuits~\cite{Yoshida,Arai} to spiking neurons~\cite{Mainen,Galan1} and ecological populations~\cite{Ranta}, where direct mutual interaction among the elements does not exist or is not appropriate to assume.
Recent studies have revealed that such synchronization generically occurs in a wide class of rhythmic systems including limit-cycle, chaotic, or stochastic oscillators, and also for various types of stochastic signals such as Gaussian, Poisson, and chaotic noise~\cite{Teramae,Goldobin,Goldobin2,Nagai,Nakao,Nakao2,Marella,Arai,Hata,Toral,Zhou,Pikovsky}. 

Efficiency of the stochastic synchronization is usually quantified by the Lyapunov exponent averaged over noise, which measures the mean exponential growth (or decay) rate of small differences between the oscillator states subjected to the common noise.
For a limit-cycle oscillator undergoing regular periodic oscillations, the Lyapunov exponent can be calculated from the phase response curve (PRC)~\cite{Winfree0,Winfree1,Kuramoto,Holmes,Goldbin3}, which is a fundamental quantity that characterizes the oscillator dynamics and which has been measured experimentally in many rhythmic elements~\cite{Khalsa,Galan2,Tateno,Robinson,Nesse}.  It can be shown that for limit-cycle oscillators driven by weak Gaussian or Poisson noise, the Lyapunov exponent is always negative irrespective of the precise shape of the PRC, ensuring that synchronization always takes place~\cite{Nakao, Hata, Arai, Pikovsky}.

What PRC shape yields the best synchronization?  For weak Gaussian driving noise, Abouzeid and Ermentrout~\cite{Abouzeid} obtained the optimal functional shape of the PRC by minimizing the Lyapunov exponent with constraints on its amplitude and smoothness, which was nearly sinusoidal with a pair of positive and negative lobes (called Type-II, a normal form of the PRC near the Hopf bifurcation~\cite{Holmes}).  However, the optimal functional shapes may differ for other driving signals.

Here we consider Poisson random impulsive signals with low frequency, which can also induce synchronization of limit cycles~\cite{Nakao,Arai,Hata,Pikovsky}.  When the intensity of the impulse is weak, the linear Gaussian approximation holds and the optimal PRC can be shown to be sinusoidal, but for stronger impulses, the optimal solution may take different shapes.
By using the shooting method~\cite{Numerical} to numerically solve the Euler-Lagrange equation~\cite{Goldstein} associated with the minimization of the Lyapunov exponent, we show that the optimal PRC gradually deviates from the sinusoid and approaches a sawtooth as the constraint on its squared amplitude is varied.  Correspondingly, the Lyapunov exponent becomes more negative and tends to diverge.
Our result implies the importance of nonlinearity in the phase response and may provide insights into real-world oscillators such as spiking neurons.

This article is organized as follows: In Sec. II, some basic facts on synchronization of limit cycles by common Poisson noise are presented.  In Sec. III, we solve the optimization problem and show the gradual transition of the optimal solution between sinusoidal and sawtoothed shapes.  Section IV summarizes the article with discussions on possible relevance of the results to phase response curves of spiking neurons. 
The appendix gives details on wavenumber, symmetry, and phase-plane behavior of the optimal solutions. We also show the optimal PRCs for stochastic desynchronzation.

%%%%%%%%%%%%%%%%%%%%%%%%%%%%%%%%%%%%%%%%%%%%%%%%%%%%%%%%%%%%%%%%%%

\section{Synchronization by common Poisson noise}

%%%%%%%%%%%%%%%%%%%%%%%%%%%%%%%%%%%%%%%%%%%%%%%%%%%%%%%%%%%%%%%%%%

\subsection{Poisson-driven oscillators}

A pair of non-interacting identical limit-cycle oscillators driven by common Poisson noise can be described by the following phase model~\cite{Nakao,Arai}:
\begin{align}
\dot{\theta_{1}}(t) &= \omega + \sum_{n=1}^{N(t)} G(\theta_{1}, c_{n}) \delta(t-t_{n}), \cr
\dot{\theta_{2}}(t) &= \omega + \sum_{n=1}^{N(t)} G(\theta_{2}, c_{n}) \delta(t-t_{n}),
\label{Eq1}
\end{align}
under the assumption that the inter-impulse intervals are sufficiently large such that the oscillator orbit perturbed by an impulse relaxes back to the original limit cycle before receiving the next impulse.
Here, $\theta_{1,2} \in [0,1)$ are phase variables of the oscillators, $\omega$ is their natural frequency, $N(t)$ is a Poisson process of rate $\lambda$, $\{t_{1}, t_{2}, \cdots\}$ are arrival times of the Poisson impulses, $\{c_{1}, c_{2}, \dots\}$ are intensities of the impulses (including negative values representing opposite directions) independently drawn from an identical probability density function (PDF) $P(c)$, and $G(\theta, c)$ is the PRC of the oscillators.

The PRC $G(\theta, c)$ quantifies the asymptotic phase difference of the orbit that is perturbed at phase $\theta$ by an impulse of intensity $c$ from the unperturbed orbit~\cite{Winfree1}.
We assume that the PRC $G(\theta, c)$ is a sufficiently smooth function with continuous derivatives $G'(\theta, c) = \partial G(\theta, c) / \partial \theta$, $G''(\theta, c) = \partial^{2} G(\theta, c) / \partial \theta^{2}$, $\cdots$, all of which are periodic in $\theta$, i.e., $G(\theta+1, c) = G(\theta)$, $G'(\theta+1, c) = G'(\theta, c)$, $\cdots$.
Equation~(\ref{Eq1}) is stochastic and should be interpreted in the Ito sense~\cite{Hanson}.  Namely, on arrival of an impulse at phase $\theta$, the phase discontinuously jumps from $\theta$ to $\theta + G(\theta, c)$~\cite{Arai}.

%%%%%%%%%%%%%%%%%%%%%%%%%%%%%%%%%%%%%%%%%%%%%%%%%%%%%%%%%%%%%%%%%%

\subsection{Lyapunov exponent}

In Refs.~\cite{Nakao,Arai}, the phase equation~(\ref{Eq1}) is derived from general limit-cycle models by the phase reduction method~\cite{Winfree0,Winfree1,Kuramoto}.  The Lyapunov exponent $\Lambda$, which quantifies the exponential growth rate of small phase differences between the oscillators $\Delta \theta(t) = \theta_{1}(t) - \theta_{2}(t)$, is given in terms of the PRC as
\begin{equation}
\Lambda = \lambda \int_0^1 d\theta P(\theta) \int dc P(c) \ln \left| 1+G'(\theta, c) \right |,
\end{equation}
where $P(\theta)$ is a stationary PDF of the phase $\theta$ given by a stationary solution of the Frobenius-Perron equation corresponding to Eq.~(\ref{Eq1})~\cite{Nakao,Arai,Hata}.
The phase difference $|\Delta \theta(t)|$ grows as
$| \Delta \theta(t) | \simeq | \Delta \theta(0) | \exp( \Lambda t )$
when it is small, so that the two oscillators tend to synchronize if the Lyapunov exponent $\Lambda$ is negative.

We assume that the impulses are sparse, i.e., the Poisson rate $\lambda$ is small.
It can then be shown that the stationary PDF of the phase $\theta$ can be approximated as
$P(\theta)=1+O(\lambda/\omega)$,
so that we may put $P(\theta)=1$ when $\lambda$ is small enough. Thus, the Lyapunov exponent is approximately given by~\cite{Nakao,Arai}
\begin{equation}
\Lambda = \lambda \int_0^1 d\theta \int dc P(c) \ln \left| 1+G'(\theta, c) \right |.
\label{Lyapunov0}
\end{equation}
Moreover, for a sufficiently smooth PRC satisfying $G'(\theta,c) > -1$, Eq.~(\ref{Lyapunov0}) can be bounded from above as
\begin{align}
\Lambda \leq & \lambda \int dc P(c) \int_{0}^{1} d\theta G'(\theta, c)\\
 = & \lambda \int dc P(c) [ G(1, c) - G(0, c) ] = 0
\end{align}
by using the inequality $\ln(1+x) \leq x$ and the periodicity of the PRC, so that $\Lambda$ is always negative (equality holds only for non-physical constant PRCs).
Thus, the two oscillators subjected to weak common Poisson noise always tend to synchronize.  Hereafter, we try to find the optimal PRC that gives the most negative Lyapunov exponent.

Note that when $\lambda$ is not sufficiently small, we may consider perturbation expansion of the stationary PDF from the uniform distribution like $P(\theta) = 1 + (\lambda/\omega) P_1(\theta) + \left ( \lambda/\omega \right )^2 P_2(\theta) + \cdots$ to calculate higher-order corrections for the Lyapunov exponent, as performed in \cite{Nakao,Arai,Abouzeid}.  For simplicity, we focus only on the case with sufficiently small $\lambda$ in the present study.
%%%%%%%%%%%%%%%%%%%%%%%%%%%%%%%%%%%%%%%%%%%%%%%%%%%%%%%%%%%%%%%%%%

\subsection{Linear Gaussian approximation}

When the derivative of the PRC $G'(\theta, c)$ is sufficiently small, we may expand Eq.~(\ref{Lyapunov0}) as
\begin{align}
\Lambda
&= \lambda \int_{0}^{1} d\theta \int dc P(c) \left( G'(\theta,c) - \frac{G'(\theta,c)^{2}}{2} + \cdots \right) \cr
&\simeq - \frac{\lambda}{2} \int dc P(c) \int_{0}^{1} d\theta G'(\theta, c)^{2},
\end{align}
where we used the periodicity of the PRC.  Also, if the impulse intensity $c$ is sufficiently weak, the PRC $G(\theta, c)$ can be linearly approximated by using the phase sensitivity function $Z(\theta)$, which gives the linear response coefficient of the phase to infinitesimal perturbations~\cite{Winfree0,Winfree1,Kuramoto}, as
\begin{align}
G(\theta, c) = c Z(\theta),
\end{align}
so that $\Lambda$ can be approximated as
\begin{align}
\Lambda &\simeq - \frac{\lambda \langle c^{2} \rangle}{2} \int_{0}^{1} d\theta Z'(\theta)^{2} \ \left( \leq 0 \right),
\label{Lyapunov1}
\end{align}
where $\langle c^{2} \rangle = \int P(c) c^{2} dc$.
This expression coincides with the Lyapunov exponent of the phase oscillators driven by weak common Gaussian-white noise~\cite{Teramae,Goldobin,Goldobin2}.
The same result can also be derived more rigorously as the diffusion limit of the Poisson noise in which the impulses tend to be weak ($c \to 0$) and frequent ($\lambda \to \infty$) while keeping $\lambda \langle c^{2} \rangle$ constant and small~\cite{Arai}.
In this diffusion limit, $P(\theta)$ can be approximated as $P(\theta) = 1 + O\left(\lambda \langle c^{2} \rangle/\omega \right)$~\cite{Abouzeid}.
Therefore, fixing the average impulse intensity small such that $\lambda \langle c^{2} \rangle \ll \omega$ is satisfied, $P(\theta)\simeq 1$ holds even for high-frequency impulses with large $\lambda$.

%%%%%%%%%%%%%%%%%%%%%%%%%%%%%%%%%%%%%%%%%%%%%%%%%%%%%%%%%%%%%%%%%%

\section{Optimal phase response curves}

%%%%%%%%%%%%%%%%%%%%%%%%%%%%%%%%%%%%%%%%%%%%%%%%%%%%%%%%%%%%%%%%%%

\subsection{Euler-Lagrange equation}

The Lyapunov exponent $\Lambda$ is a functional of the PRC $G(\theta, c)$ or the phase sensitivity function $Z(\theta)$ as given in Eq.~(\ref{Lyapunov0}) or Eq.~(\ref{Lyapunov1}).  We try to obtain the optimal shape of $G(\theta, c)$ or $Z(\theta)$ for synchronization by minimizing $\Lambda$ with appropriate constraints.
Let us omit the dependence of the PRC $G(\theta,c)$ on $c$ for the moment.  We try to find the minimum of the action~\cite{Goldstein}
\begin{align}
S[G] = & \Lambda[G] + \mu J[G] + \nu K[G]\cr
= & \int_{0}^{1} L(G(\theta), G'(\theta), G''(\theta)) d\theta,
\end{align}
where $\Lambda[G]$ is the Lyapunov exponent, $J[G]$ and $K[G]$ are two independent constraints on the PRC and its derivatives (we consider up to the 2nd order), $\mu$ and $\nu$ are Lagrange multiplier{s}, and $L(G(\theta), G'(\theta), G''(\theta))$ is a Lagrangian.
The corresponding Euler-Lagrange equation is given by
\begin{align}
\frac{d^2}{d\theta^2} \frac{\partial L}{\partial G''} - \frac{d}{d\theta} \frac{\partial L}{\partial G'} + \frac{\partial L}{\partial G} = 0,
\label{EL}
\end{align}
where the periodicity of the PRC and the derivative, $G(\theta+1) = G(\theta)$ and $G'(\theta+1) = G'(\theta)$, are used to eliminate the surface terms.
When we consider optimization of Eq.~(\ref{Lyapunov1}), the PRC $G(\theta)$ in the above equations is replaced by the phase sensitivity function $Z(\theta)$.

%%%%%%%%%%%%%%%%%%%%%%%%%%%%%%%%%%%%%%%%%%%%%%%%%%%%%%%%%%%%%%%%%%

\subsection{Linear Gaussian approximation}

We here briefly explain the optimal $Z(\theta)$ under linear approximation.  See Abouzeid and Ermentrout~\cite{Abouzeid} for a detailed analysis with various constraints.
From Eq.~(\ref{Lyapunov1}), the Lyapunov exponent of the oscillator in this case is given by
\begin{equation}
\Lambda_0[Z] = -\frac{D}{2} \int_0^1  d\theta Z'(\theta)^2,
\end{equation}
where $D = \lambda \langle c^{2} \rangle$ corresponds to the intensity or variance of the driving noise.
We calculate the optimal shape of $Z(\theta)$ by minimizing $\Lambda_0[Z]$ under the following constraints:
\begin{align}
J_0[Z] = & \int_0^1 Z(\theta)^2 d\theta - B_{0} = 0,\label{eq01}\\
K_0[Z] = & \int_0^1 Z''(\theta)^2 d\theta - C_{0} = 0.\label{eq01_2}
\end{align}
The first constraint Eq.~(\ref{eq01}) fixes the squared amplitude of $Z(\theta)$ to be $B_{0}$, which excludes the possibility of non-physical divergent $Z(\theta)$ yielding arbitrarily negative Lyapunov exponents.
The second constraint Eq.~(\ref{eq01_2}) with parameter $C_{0}$ restricts the overall smoothness of $Z(\theta)$.  In most realistic finite-dimensional limit-cycle oscillators, the first Fourier mode dominates the phase sensitivity function $Z(\theta)$, reflecting the circular geometry of the limit cycle orbit in the phase space.
We thus introduce the constraint Eq.~(\ref{eq01_2}) to avoid rapid oscillations and choose physically natural PRCs, similarly to Ref.~\cite{Abouzeid}.

Introducing Lagrange multipliers $\mu_0$ and $\nu_0$, the action to be minimized is given by
\begin{align}
S_0[Z] =& \Lambda_0[Z] +\mu_0 J_0[Z] + \nu_0 K_0[Z]\cr
= & \int_0^1 \left\{ -\frac{D}{2} Z'(\theta)^2 +\mu_0 \left ( Z(\theta)^2 - B_{0} \right )+\nu_0 \left ( Z''(\theta)^2 - C_{0} \right ) \right \} d\theta\cr
= & \int_0^1 L_0(Z(\theta), Z'(\theta), Z''(\theta)) d\theta.
\label{eq_EL1}
\end{align}
The Euler-Lagrange equation determining the optimal $Z(\theta)$ is given by
\begin{equation}
2\nu_0 Z^{(4)}(\theta)+DZ''(\theta)+2\mu_0 Z(\theta)=0,
\label{eq02}
\end{equation}
where $Z^{(4)}$ denotes the 4th derivative of $Z$.
When $\mu_0>0$ and $\nu_0>0$, we obtain a general solution that satisfies the periodic boundary condition $Z(\theta) = Z(\theta+1)$ as
\begin{equation}
Z(\theta) = \alpha \sin \left( \sqrt{\frac{D\pm \sqrt{D^2-16\mu_0\nu_0}}{4\nu_0}} \theta + \beta \right),
\end{equation}
where $\alpha$ and $\beta$ are constants.
Due to the periodicity $Z(\theta) = Z(\theta+1)$, the coefficient of $\theta$ should be quantized as
\begin{equation}
\sqrt{\frac{D\pm \sqrt{D^2-16\mu_0\nu_0}}{4\nu_0}} = 2 \pi n,
\label{aeq17}
\end{equation}
where $n$ is an integer number.  The constant $\alpha$ is determined from the first constraint Eq.~(\ref{eq01}) as
\begin{align}
  \int_0^1 Z(\theta)^2 d\theta = \frac{\alpha^2}{2} = B_{0},
\end{align}
namely, $\alpha = \sqrt{2 B_{0}}$.
The constant $\beta$ is determined from the boundary conditions for $Z(\theta)$.  Without losing generality, we can assume that $Z(0) = 0$ and $Z'(0) > 0$, which yields $\beta = 0$.
The second constraint Eq.~(\ref{eq02}) gives
\begin{align}
  \int_0^1 Z''(\theta)^2 d\theta = B_{0} \left ( \frac{D\pm \sqrt{D^2-16\mu_0\nu_0}}{4\nu_0}\right )^2 = C_{0}.
\label{aeq19}
\end{align}
Equations~(\ref{aeq17}) and (\ref{aeq19}) give the relation between Lagrange multipliers ($\mu_0$, $\nu_0$) and the parameters ($B_{0}$, $C_{0}$).
In the following, we will control the Lagrange multipliers to find optimal solutions with given squared amplitude and overall smoothness.

The optimal phase sensitivity function is thus given by
\begin{equation}
Z(\theta) = \sqrt{2 B_{0}} \sin ( 2 \pi n\theta ),
\label{eq3}
\end{equation}
which is always sinusoidal regardless of the constraint parameters.  The corresponding Lyapunov exponent is obtained from Eq.~(\ref{Lyapunov1}) as
\begin{equation}
\Lambda_0 = - \frac{D B_{0}}{2} n^{2},
\end{equation}
which decreases with the wavenumber $n$ without bounds.  Namely, rapidly oscillating $Z(\theta)$ can yield very small $\Lambda_{0}$ if the constraint on the smoothness of $Z(\theta)$ does not exist.
The second constraint Eq.~(\ref{eq01_2}) restricts the range of the wavenumber $n$.  In particular, when $\nu_0$ is sufficiently large, only small $n$ is allowed (See appendix). 
To obtain realistic PRCs, we thus set the parameter $\nu_0 > 0$ large enough and focus on $Z(\theta)$ with $n=1$ as well as the corresponding $G(\theta)$, namely, we look for the optimal PRC having only a single pair of positive and negative lobes (Type-II) that oscillates only once in $\theta \in [0,1)$ and crosses the $\theta$-axis exactly twice, which is typical of realistic limit-cycle oscillators.

%%%%%%%%%%%%%%%%%%%%%%%%%%%%%%%%%%%%%%%%%%%%%%%%%%%%%%%%%%%%%%%%%%

\subsection{Poisson impulses}

What is the optimal shape of the PRC when the oscillators are driven by common Poisson noise?
As we saw, when the applied impulse is sufficiently weak and the amplitude of the PRC is small enough, linear Gaussian approximation holds and the optimal PRC is sinusoidal.
But linear approximation may not be valid when the impulse intensity is increased~\cite{Nakao}.
On the other hand, if no constraint is imposed on the PRC, an obvious optimal solution is a sawtooth, consisting of a straight line of slope $-1$ and 
a sharp jump to satisfy the periodic boundary conditions.
The corresponding Lyapunov exponent diverges to $-\infty$, because a single impulse can already synchronize the oscillators by instantaneously reseting their phases to the same value.
However, if the impulse is not sufficiently strong to kick the oscillator, such a simple solution is impossible.
How does the optimal PRC behave in between the two limiting situations?

In the following, we focus on two simple cases in which the oscillators are driven by (i) excitatory impulses with a constant intensity (all impulses take the same intensity $c$), and (ii) both excitatory and inhibitory impulses (the impulses take either $c=a$ or $c=-a$ with equal probability).
We examine how the optimal PRC deviates from the sinusoid and eventually approaches the trivial sawtooth shape as the constraint on the squared amplitude of the PRC is increased.

%%%%%%%%%%%%%%%%%%%%%%%%%%%%%%%%%%%%%%%%%%%%%%%%%%%%%%%%%%%%%%%%%%

\subsubsection{Excitatory impulses}

We assume that the impulse intensity $c$ always takes the same value and simply denote the PRC corresponding to this value as $G(\theta)$.  The Lyapunov exponent is given by
\begin{align}
\Lambda_1[G]
= & \lambda \int_0^1  \ln \left| 1+G'(\theta) \right | d\theta.
\end{align}
We minimize $\Lambda_1[G]$ under the constraints on squared amplitude and overall smoothness of $G$,
\begin{align}
J[G] = & \int_0^1 G(\theta)^2 d\theta -  B = 0,\label{const1}\\
K[G] = & \int_0^1 G''(\theta)^2 d\theta -  C = 0,\label{const2}
\end{align}
and examine the dependence of the optimal PRC on the parameter $B$ that determines the squared amplitude while fixing $C$ small enough (actually taking the value of $\nu$ appropriately large) such that the PRC keeps a given level of smoothness.

Introducing Lagrange multipliers $\mu$ and $\nu$, the action to be minimized is given as
\begin{align}
S_{1}[G] = & \Lambda_1[G] + \mu J[G] + \nu K[G]\cr
 = & \int_0^1 \left\{ \lambda \ln \left | 1+G'(\theta) \right | + \mu\left ( G(\theta)^2-B\right ) + \nu\left ( G''(\theta)^2-C\right ) \right \} d\theta \cr
= &\int_0^1 L_1(G, G', G'') d\theta.
\label{eq_EL2}
\end{align}
The optimal PRC $G(\theta)$ is determined by the Euler-Lagrange equation
\begin{align}
\frac{d^2}{d\theta^2}\frac{\partial L_1}{\partial G''} - \frac{d}{d\theta}\frac{\partial L_1}{\partial G'} + \frac{\partial L_1}{\partial G} = 0,
\end{align}
which gives
\begin{align}
\nu G^{(4)} + \frac{\lambda}{2}\frac{G''}{(1+G')^{2}} + \mu G = 0,\label{eq22}
\end{align}
where $G^{(4)}$ denotes the fourth derivative of $G$.

If the squared amplitude of the PRC $B$ is sufficiently small, linear approximation for the PRC should hold, i.e., $G(\theta) = \epsilon Z(\theta)$ where $\epsilon \ (\propto \sqrt{B})$ is a small constant. 
The constraints Eqs.~(\ref{const1}) and (\ref{const2}) become equivalent to Eqs.~(\ref{eq01}) and (\ref{eq01_2}) under the linear approximation by rescaling the multipliers as $\mu = \mu_0 / \epsilon^2$ and $\nu = \nu_0 / \epsilon^2$.
%%
%%\begin{align}
%%\mu \int_0^1 G^2 d\theta \sim & \frac{\mu_0}{\epsilon^2} \int_0^1 \epsilon^2 G^2 d\theta =  \mu_0 \int_0^1 Z^2 d\theta\\
%%\nu \int_0^1 G''^2 d\theta \sim & \frac{\nu_0}{\epsilon^2} \int_0^1 \epsilon^2 G''^2 d\theta =  \nu_0 \int_0^1 Z''^2 d\theta.
%%\end{align}
%%}
%%
Substituting these into Eq.~(\ref{eq22}), we obtain
\begin{equation}
\nu_0 Z^{(4)} + \frac{\lambda \epsilon^2}{2}\frac{Z''}{(1+\epsilon Z')^{2}} + \mu_0 Z = 0,
\end{equation}
and taking the $\epsilon \to 0$ limit with $D=\lambda\epsilon^2$ fixed, we obtain the Euler-Lagrange equation~(\ref{eq02}) for weak Gaussian noise and thus yields sinusoidal $Z(\theta)$ and $G(\theta)$ as the optimal solution.
On the other hand, if we ignore the constraint Eq.~(\ref{const1}), $G(\theta) = - \theta + const.$ is a trivial solution to Eq.~(\ref{eq22}), which gives a sawtooth.
Thus, when the squared amplitude of $G(\theta)$ is controlled, mutation of the optimal PRC between the two limiting shapes is expected.

To confirm this, we numerically calculate a family of solutions to Eq.~(\ref{eq22}) using the shooting method~\cite{Numerical}.  Namely, we numerically integrate Eq.~(\ref{eq22}) by the Runge-Kutta method with adaptive time grids from $\theta=0$ to $\theta=1$ and find appropriate initial conditions $G(0)$, $G'(0)$, $G''(0)$ and $G'''(0)$ satisfying the periodic boundary conditions at $\theta=0$ and $\theta=1$.
We vary the Lagrange multiplier $\mu > 0$, obtain the corresponding optimal PRC, and check if its squared amplitude was equal to the constraint $B$.
Solutions to Eq.~(\ref{eq22}) exist also for $\mu < 0$, but they {\it maximize} the Lyapunov exponent rather than minimize it, and thus are optimal not for synchronization but for {\it desynchronization} (see Appendix).
It can be shown that large values of $\nu$ lead to small wavenumber (long wavelength) solutions (see Appendix).
We fix the multiplier $\nu$ at $\nu = 10^{-5}$, which is large enough,  to choose non-trivial solutions that cross the $\theta$-axis exactly twice in $[0,1)$ corresponding to the $n=1$ case in Eq.~(\ref{eq3}).
No periodic solutions are found when $\nu<0$.
Properties of the optimal solution can be well understood by approximate phase-plane analysis as explained in Appendix.

Figure~\ref{fig01}(a) shows the results, where the optimal solutions are wrapped within the range $[-0.5, 0.5)$ by taking modulo $1$.
All solutions lay within the plotted region, and no other solutions outside of this region are found.
The solutions are symmetric with respect to $\theta = 0.5$ reflecting the symmetry of the Euler-Lagrange equation~(\ref{eq22}) (see Appendix).
As expected, we see that the optimal PRC is almost sinusoidal when the parameter $B$ is small.  As $B$ is increased, the optimal PRC gradually deviates from the sinusoid and approaches a symmetric sawtooth limit (which gives $B=1/12$).
Correspondingly, the Lyapunov exponent $\Lambda_{1}$ plotted in Fig.~\ref{fig01}(b) becomes more negative and tends to diverge, and its inverse $\tau_{1} = -1 / \Lambda_{1}$, which gives characteristic time for the stochastic synchronization, gradually decreases to zero as shown in Fig.~\ref{fig01}(c).

The optimality of the obtained PRC can be clearly demonstrated by numerical simulations.  Figure~\ref{fig02} shows realizations of the stochastic synchronization processes with the optimal and suboptimal (sinusoidal) PRCs.  We see that the stochastic synchronization occurs much faster when the optimal PRC is used.

%% fig 1 %%

\begin{figure}[tbhp]
\begin{center}
\includegraphics[width=0.6\hsize]{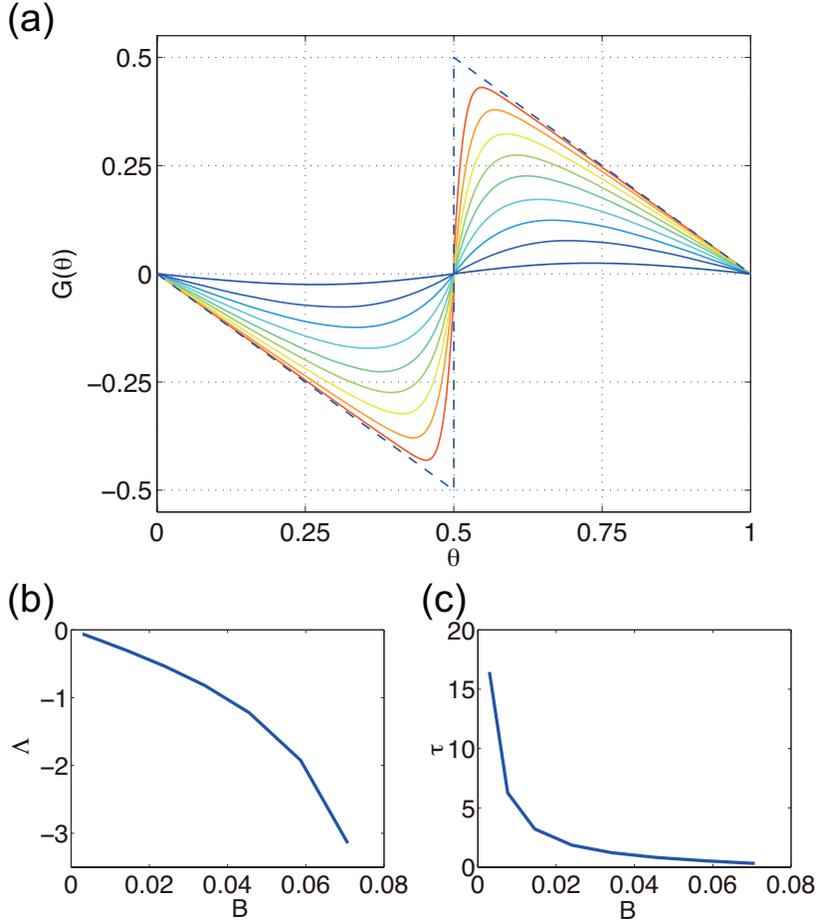}
\caption{
(Color online) (a) Numerical solutions of the Euler-Lagrange equation~(\ref{eq22}) obtained by the shooting method.  The dashed line plots the limiting sawtooth solution.  The solid curves are non-trivial solutions for various values of the squared amplitude ranging from $B=2.98\times10^{-4}$ to $7.07\times 10^{-2}$.  (b) Dependence of the Lyapunov exponent $\Lambda_1$ on $B$.  (c) Dependence of the characteristic synchronization time $\tau_1 = - 1 / \Lambda_{1}$ on $B$. 
}
\label{fig01}
\end{center}
\end{figure}

%% fig 2 %%

\begin{figure}[tbhp]
\begin{center}
\includegraphics[width=0.6\hsize]{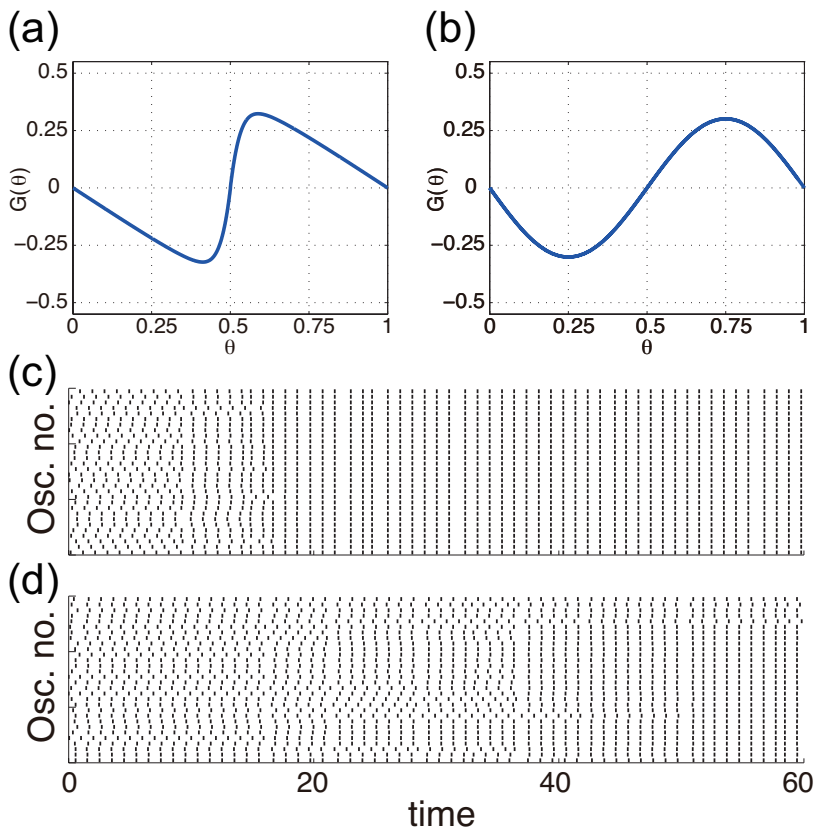}
\caption{
(Color online) Numerical realizations of the stochastic synchronization processes with the optimal and suboptimal (sinusoidal) PRCs for the case of excitatory impulses.  The squared amplitude of both PRCs is set at the same value $B=0.045$.
(a) Optimal PRC (Lyapunov exponent $\Lambda_1 = -1.221$).
(b) Sinusoidal suboptimal PRC (Lyapunov exponent $\Lambda_1 = -0.025$).
(c),(d) Numerical realizations of the stochastic synchronization processes with the PRCs shown in (a),(b).}
\label{fig02}
\end{center}
\end{figure}

%%%%%%%%%%%%%%%%%%%%%%%%%%%%%%%%%%%%%%%%%%%%%%%%%%%%%%%%%%%%%%%%%%

\subsubsection{Excitatory and inhibitory impulses}

We next consider the case that the intensity of the impulses takes two values $\pm a$ with equal probability, namely, $P(c) = \left [ \delta(c-a) + \delta(c+a) \right ] / 2$.  The Lyapunov exponent is
\begin{align}
\Lambda = & \lambda \int_0^1 \frac{1}{2} \left \{ \delta(c-a) + \delta(c+a) \right \} \ln \left| 1+G'(\theta, c) \right | d\theta dc\cr
= &  \frac{\lambda}{2}\int_0^1  \ln \left| \left(1+G'(\theta, a)\right )\left(1+G'(\theta, -a)\right ) \right | d\theta.
\end{align}
For simplicity, we seek for symmetric PRCs that satisfy $G(\theta, -a) = -G(\theta, a)$.  This condition should be always satisfied if $a$ is sufficiently small, because the PRC can be linearly approximated as $G(\theta, c) = c Z(\theta)$.
The existence of the diffusion limit is also ensured with this condition~\cite{Arai}.
Note that, for stronger impulses, the PRC generally becomes asymmetric and does not satisfy the above condition. We here focus only on the symmetric case for simplicity.

The Lyapunov exponent is then given by
\begin{equation}
\Lambda_2[G] =  \frac{\lambda}{2} \int_0^1 \ln \left| 1-G'(\theta)^2 \right | d\theta
\end{equation}
with the abbreviation $G(\theta) = G(\theta, a)$.  We minimize $\Lambda_2[G]$ under the constraints (\ref{const1}) and (\ref{const2}).
Introducing Lagrange multipliers $\mu$ and $\nu$, we obtain the action
\begin{align}
S_{2}[G] = & \Lambda_2[G] + \mu J[G] + \nu K[G]\cr
= & \int_0^1 \left \{ \frac{\lambda}{2} \ln \left | 1-G'(\theta)^2 \right | + \mu\left ( G(\theta)^2-B\right )+\nu \left ( G''(\theta)^2-C\right )\right \} d\theta \cr
= &\int_0^1 L_2(G, G', G'') d\theta,
\end{align}
and the associated Euler-Lagrange equation
\begin{align}
\nu G^{(4)}(\theta) + \frac{\lambda}{2}\frac{G''(1+G'^2)}{(1-G'^2)^2} + \mu G = 0.
\label{eq31}
\end{align}
If the squared amplitude $B$ of the PRC is sufficiently small, we can rewrite Eq.~(\ref{eq31}) using the linear approximation of the PRC with rescaled multipliers, $G(\theta) = G(\theta, a) = a Z(\theta)$, $\mu =\mu_0 / a^2$ and $\nu = \nu_0 / a^2$, as
\begin{align}
\nu Z^{(4)}(\theta) + \frac{\lambda a^2}{2} \frac{Z''(1 + a^2 Z'^2)}{(1 - a^2 Z'^2)^{2}} + \mu Z = 0.
\end{align}
Taking the diffusion limit, i.e., $a\rightarrow 0$ and $\lambda \to \infty$ with $D = \lambda a^2$ fixed, the Euler-Lagrange equation~(\ref{eq02}) under the linear Gaussian approximation is derived.
Therefore, we obtain a sinusoidal $Z(\theta)$ and hence $G(\theta)$ as the optimal solution for small $B$.
On the other hand, if we ignore the constraint, Eq.~(\ref{eq31}) has the obvious solution $G(\theta) = -\theta$ as before.  In the present case, additionally, $G(\theta) = \theta$ is also an optimal solution because $G(\theta, -a) = -G(\theta, a)$.

\begin{figure}[tbhp]
  \begin{center}
\includegraphics[width=0.6\hsize]{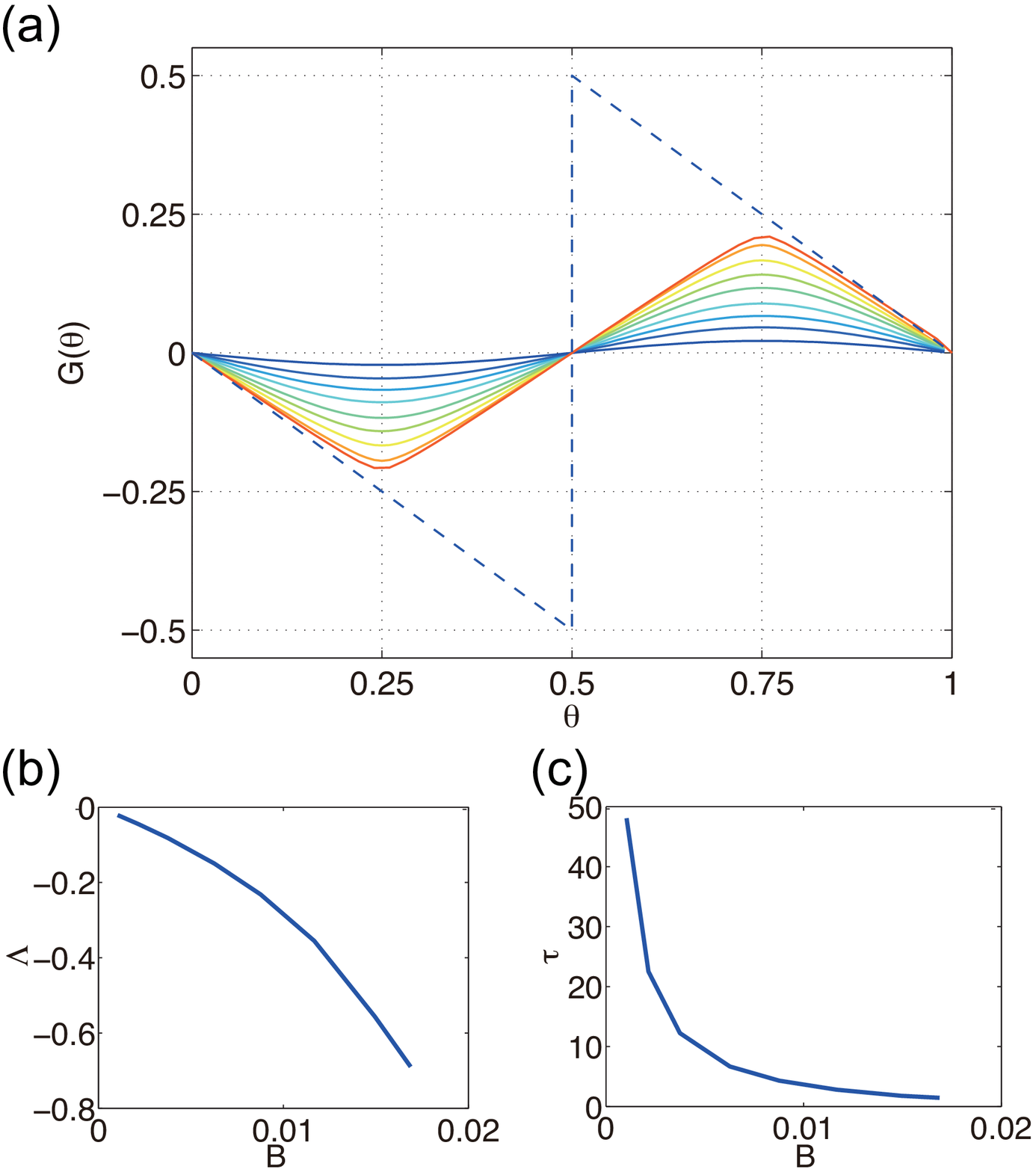}
   \caption{(Color online) Numerical solutions of the Euler-Langrange equation~(\ref{eq31}) obtained by the shooting method.
   (a) The dashed line plots the trivial solution.  The solid curves are non-trivial solutions for various values of the squared amplitude ranging from $B=1.04\times10^{-3}$ to $1.70\times 10^{-2}$.  (b) Dependence of the Lyapunov exponent $\Lambda_2$ on $B$.  (c) Dependence of the the characteristic synchronization time $\tau_2 = - 1 / \Lambda_{2}$ on $B$. 
   }
   \label{fig03}
  \end{center}
\end{figure}

\begin{figure}[tbhp]
\begin{center}
\includegraphics[width=0.6\hsize]{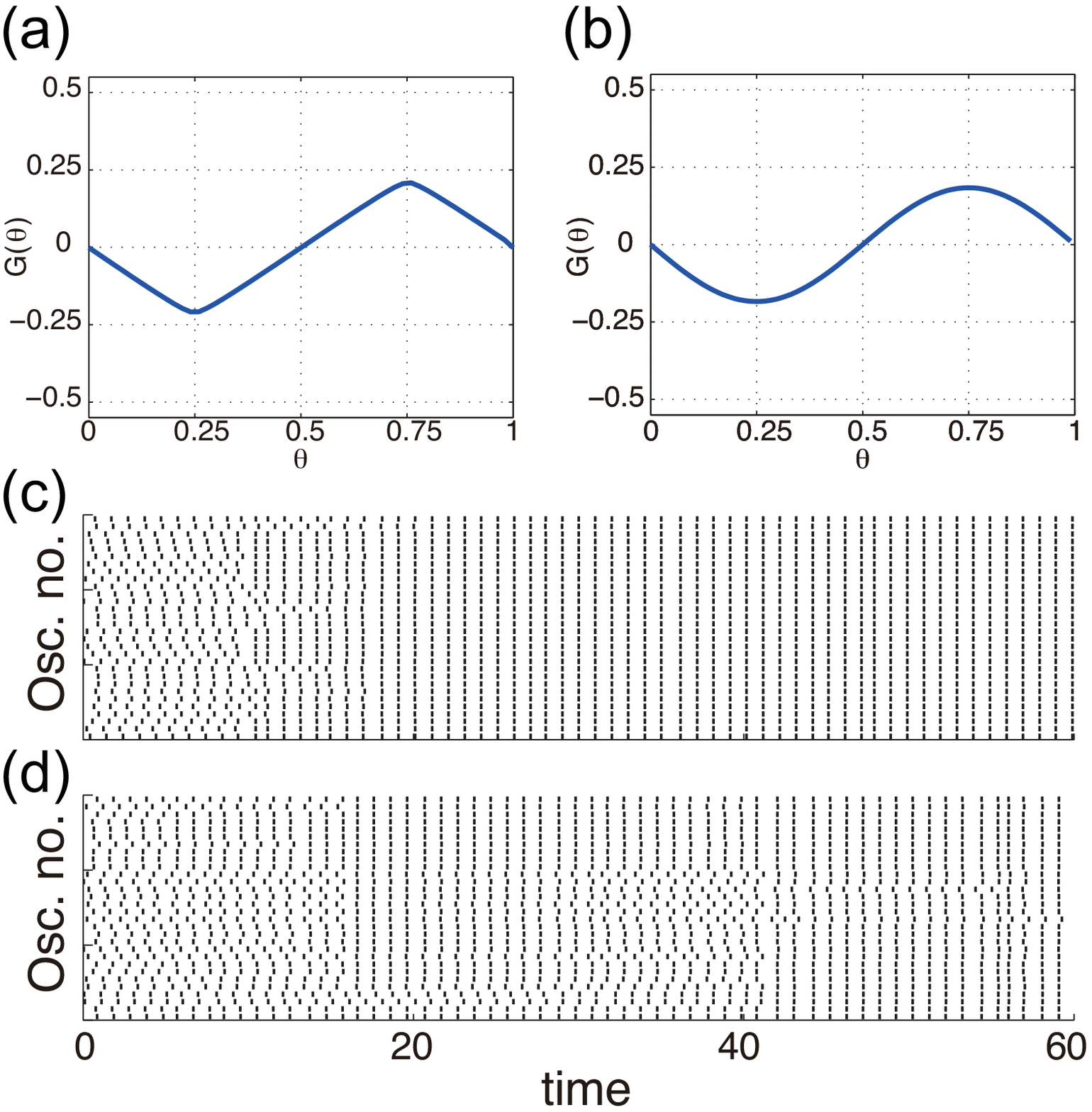}
\caption{
(Color online) Stochastic synchronization processes with the optimal and suboptimal PRCs for the case of both excitatory and inhibitory impulses.  The squared amplitude of both PRCs is set as the same value $B=0.017$.  (a) Optimal PRC (Lyapunov exponent $\Lambda_1 = -0.690$).  (b) Sinusoidal PRC (Lyapunov exponent $\Lambda_1 = -0.544$).  (c), (d) Numerical realizations of the stochastic synchronization processes with the PRCs shown in (a), (b).
}
\label{fig04}
\end{center}
\end{figure}

Using the numerical shooting method, we obtain a family of optimal solutions to Eq.~(\ref{eq31}) as plotted in Fig.~\ref{fig03}(a).  As in the previous case, the multiplier $\nu$ is fixed at $10^{-5}$, which is large enough to yield smooth PRCs.
Unlike the previous case, no solution with period $1$ exists when $\mu<0$.
As the parameter $B$ increases, the optimal PRC gradually deviates from the sinusoid.  In this case, the PRC approaches a double sawtooth, in contrast to the single sawtooth that we obtained previously, reflecting the symmetry assumption.
The Lyapunov exponent $\Lambda_{2}$ becomes more negative and tends to diverge, and the characteristic synchronization time $\tau_{2}$ decreases to zero as shown in Figs.~\ref{fig03}(b) and (c).
The optimality can be demonstrated by numerical simulation as shown in Fig.~\ref{fig04}. 

%% fig 3 %%

%% fig 4 %%

%%%%%%%%%%%%%%%%%%%%%%%%%%%%%%%%%%%%%%%%%%%%%%%%%%%%%%%%%%%%%%%%%%

\section{Discussion}

We considered the optimization problem of the PRC for synchronization of limit-cycles oscillators by common Poisson noise and observed a crossover of the optimal PRC from a sinusoid to a sawtooth by increasing its squared amplitude.
Now we take some time to stress the importance of considering nonlinear PRCs.  The phase sensitivity function $Z(\theta)$ quantifies the linear response property of the oscillator phase to infinitesimal perturbations, which is determined by the local phase-space structure of the oscillator near the limit-cycle orbit~\cite{Winfree0,Winfree1,Kuramoto}.
In contrast, the PRC can reflect nonlinear dynamics of the oscillator away from the limit-cycle orbit by finite distances, providing more detailed information.  
Also, in many experiments, applied perturbations to the oscillator are not always sufficiently small and nonlinear effects can become important.
In the present study, we considered only two simple types of driving impulses, i.e., (i) excitatory and (ii) excitatory and inhibitory impulses, and also assumed symmetry of the PRCs in the latter case.  More general types of driving impulses and asymmetric PRCs can be considered within the same framework, though they are beyond the scope of the present study.
For example, it would be interesting to seek for the optimal family of PRCs for a given distribution of the impulse intensity $c$ by making additional assumptions on the $c$-dependence of the PRC $G(\theta, c)$.

Are there examples of the optimal PRC in nature?
In neurophysiology, the PRCs of periodically spiking cells have been recorded in many experiments~\cite{Galan2,Tateno,Robinson,Nesse}.  For example, Tateno and Robinson~\cite{Tateno} calculated the PRCs of periodically spiking interneurons from monkey somatosensory cortex and examined their dependence on the intensity of applied perturbations.  As the intensity increases, the PRC changes its shape from sinusoidal to sawtoothed (Figs.~4 and 5 in Ref.~\cite{Tateno}).  This dependence of the PRC on the applied signal intensity resembles the gradual transition that we obtained in Fig.~\ref{fig01}.  The authors also found that the sawtooth-like PRCs lead to faster synchronization of the neurons~\cite{Robinson}.
Nesse and Clark~\cite{Nesse} calculated the PRC of photoreceptor cells from marine invertebrate {\it Hermissenda} and revealed  noticeable linear dependence of the PRC on $\theta$ (Figure~5 in Ref.~\cite{Nesse}).  The authors suggested that the reset effect of such a PRC may be helpful for network information processing.

Stochastic synchrony can be a mechanism for long-range synchronization of gamma oscillations in the cortex~\cite{Galan1}.
Because the thalamus is at the center of the brain and communicates with all cortical regions, it is a good candidate to provide common input to areas of the cortex that are far apart and not directly connected. This shared thalamic drive represents a straightforward mechanism to mediate synchrony between these areas, and the PRC of neurons in cortex receiving thalamic input could be optimized for this purpose.
In Ref.~\cite{Lefort}, Lefort {\it et al.} report that synaptic strength of neurons in some cortical areas have a long-tailed distribution, meaning certain synapses are much stronger than others (the amplitude of postsynaptic potentials spans over a few millivolts).  Thus, it might actually be more appropriate to consider finite-intensity impulses than weak Gaussian noise as the driving signal to the neurons.

The optimization viewpoint may give interesting insights into the understanding of biological systems, because they evolved to perform certain biological functions efficiently.  If the stochastic synchronization mechanism is used in some biological systems, their PRC may be optimized to best perform synchronization.
The sawtoothed PRCs that we obtained are not only optimal for the synchronization by common Poisson noise, but they are singular in the sense that they lead to instantaneous phase resetting of the oscillators.  Thus, it may not be surprising if such a singular shape is actually utilized in real biological systems.
This parallel between evolutionary optimization and optimization for a desired function certainly makes the interpretation of such a singular shape highly suggestive and intriguing.

{\it Acknowledgements.}  We thank H. P. C. Robinson for useful comments.  S.H. is supported by the GCOE program ``The Next Generation of Physics, Spun from Universality and Emergence'' from MEXT, Japan. H.N. thanks financial support by MEXT, Japan (grant no. 22684020). R.F.G is supported by The Mount Sinai Healthcare Foundation and The Alfred P. Sloan Foundation, USA.

%%%%%%%%%%%%%%%%%%%%%%%%%%%%%%%%%%%%%%%%%%%%%%%%%%%%%%%%%%%%%%%%%%

\section*{APPENDIX}
In this Appendix, we give detailed discussion on the dependence on the Lagrange multiplier $\nu$, symmetry, and phase-plane analysis, of the optimal solutions.  We also show the optimal PRCs for stochastic desynchronization.

%%%%%%%%%%%%%%%%%%%%%%%%%%%%%%%%%%%%%%%%%%%%%%%%%%%%%%%%%%%%%%%%%%

\subsection{Dependence of the optimal PRCs on the multipliers}

We find that if the multiplier $\nu_0$ or $\nu$ is sufficiently large, only small wavenumber (long wavelength) solutions are allowed for $Z$ or $G$.  This can be proven for the Euler-Lagrange equations~(\ref{eq02}), (\ref{eq22}) and (\ref{eq31}).

\subsubsection{Linear Gaussian approximation}

We consider a solution of the Euler-Lagrange equation (\ref{eq02}) with wavenumber $n$ and denote the corresponding Lagrange multipliers ($\mu_n, \nu_n$).  Substitution into (\ref{eq02}) yields
\begin{align}
8\pi^4 \nu_n n^2 + D \pi^{2} + \frac{\mu_n}{n^2} & = 0,
\end{align}
namely, the Lagrange multipliers scale with the wavenumber $n$ as
$\mu_n \propto n^2$ and $\nu_n \propto 1/ n^2$.
Thus, larger $\mu$ and smaller $\nu$ lead to PRCs with larger wavenumbers.  We set $\nu$ sufficiently large to obtain the $n=1$ solution in the main text.

\subsubsection{Poisson impulses}

Rescaling the phase variable as $\theta \to n \theta$, the Euler-Lagrange equaion~(\ref{eq22}) is transformed to
\begin{align}
\nu \frac{1}{n^{2}} \frac{d^{4}}{d \theta^{4}} G(n \theta) + \frac{\lambda}{2} \frac{ \frac{d^{2}}{d \theta^{2} }G(n \theta)}{( 1 + \frac{1}{n} \frac{d}{d\theta} G(n \theta) )^2}+ \mu n^{2} G(n \theta) =0.
\end{align}
Defining a rescaled PRC $G_{n}(\theta) = G(n \theta) / n$, the above equation can be cast into the same form as Eq.~(\ref{eq22}),
\begin{align}
%%\nu_{n} \frac{d^{4}}{d \theta^{4}} G_{n}(\theta) + \frac{\lambda}{2} \frac{ \frac{d^{2}}{d \theta^{2} }G(\theta)}{(1 + \frac{d}{d\theta} G_{n}(\theta))^2}+ \mu_{n} G_{n}(\theta) =0,
\nu_{n} G^{(4)}_{n}(\theta) + \frac{\lambda}{2} \frac{G^{2}(\theta)}{(1 + G'_{n}(\theta))^2} + \mu_{n} G_{n}(\theta) =0,
\label{eq_scaling}
\end{align}
where rescaled Lagrange multipliers $\mu_n = n^2 \mu$ and $\nu_n = \nu / n^2$ are introduced.
Thus, if $G(\theta)$ is a solution of Eq.~(\ref{eq22}) with multipliers $\mu$ and $\nu$, its rescaled function $G_{n}(\theta)$ is also a solution of Eq.~(\ref{eq22}) with rescaled multipliers $\mu_{n}$ and $\nu_{n}$ ($n=1, 2, \cdots$). 
This implies that larger wavenumber solutions ($n>1$) correspond to larger $\mu$ and smaller $\nu$.
As shown in Fig.~\ref{fig01}, the multiplier $\mu$ control the squared amplitude $B$ and determine the shape of the periodic solutions.
Thus, $\nu$ determines the wavenumber of the solution, which we take sufficiently large ($\nu = 10^{-5}$) to obtain the PRC corresponding to $n=1$.
Similarly, rescaling Eq.~(\ref{eq31}), we obtain
\begin{align}
\nu_n G^{(4)}_{n}(\theta) + \frac{\lambda}{2} \frac{ G^{(2)}_{n}(\theta) ( 1 + G'_{n}(\theta)^2 )}{ ( 1 - G'_{n}(\theta)^{2} )^2} + \mu_n G_{n}(\theta) =0\nonumber
\end{align}
to find $\mu_n = n^2\mu$ and $\nu_n = \nu / n^2$.  Thus, if $\nu$ is sufficiently large, the PRC takes the smallest wavenumber $n=1$.

%%%%%%%%%%%%%%%%%%%%%%%%%%%%%%%%%%%%%%%%%%%%%%%%%%%%%%%%%%%%%%%%%%

\subsection{Symmetry of the optimal solution}

From Eqs.~(\ref{eq22}) and (\ref{eq31}) with periodic boundary conditions $G(0) = G(1) = 0$, $G'(0) = G'(1)$, and $G''(0) = G''(1)$, we obtained symmetric solutions with respect to $\theta=0.5$ as shown in Figs.~\ref{fig01}(a) and \ref{fig03}(a). 
These PRCs inherit the symmetry from the Euler-Lagrange equations (or the actions to be minimized).
To see this, let us define a function $F(\theta)$ as
\begin{equation}
F(\theta) = -G(1-\theta).\label{eq_sym}
\end{equation}
The derivatives are $F'(\theta) = G'(1-\theta)$, $F''(\theta) = -G''(1-\theta)$, and $F^{(4)}(\theta) = -G^{(4)}(1-\theta)$.  Substituting into Eq.~(\ref{eq22}), 
we find that $F(\theta)$ obeys the same Euler-Lagrange equation as $G(\theta)$,
%%
%\begin{equation}
%F''(\theta) = - \frac{2\mu}{\lambda} F(\theta) \left ( 1+F'(\theta)^2 \right ),
%\label{eqA2}
%\end{equation}
\begin{align}
\nu F^{(4)}(\theta) + \frac{\lambda}{2}\frac{F''}{(1+F')^2} + \mu F = 0,
\label{eqA2}
\end{align}
with the same boundary conditions $G(0) = F(0)$, $G'(0) = F'(0)$, and $G''(0) = F''(0)$ (note that all the solutions satisfy $G''(0) = 0$).
We thus obtain $G(\theta) \equiv  F(\theta) = -G(1-\theta)$, indicating that $G(\theta)$ is symmetric with respect to $\theta = 0.5$.

Similarly, when the optimal PRC $G(\theta)$ obeys the Euler-Lagrange equation~(\ref{eq31}), we can show that $F(\theta)$ satisfies the same equation
%%
%\begin{equation}
%F''(\theta) = -\frac{2\mu}{\lambda} \frac{F(\theta) \left ( 1-F'(\theta)^2 \right )^2}{1+F'(\theta)^2}
%\end{equation}
\begin{equation}
\nu F^{(4)}(\theta) + \frac{\lambda}{2}\frac{F''(1+F'^2)}{(1-F'^2)^2} + \mu F = 0
\end{equation}
with the same boundary conditions.  Thus, $G(\theta) \equiv F(\theta) = -G(1-\theta)$ holds and $G(\theta)$ is also symmetric with respect to $\theta = 0.5$.

%%%%%%%%%%%%%%%%%%%%%%%%%%%%%%%%%%%%%%%%%%%%%%%%%%%%%%%%%%%%%%%%%%

\subsection{Phase-plane analysis}

%%%%%%%%%%%%%%%%%%%%%%%%%%%%%%%%%%%%%%%%%%%%%%%%%%%%%%%%%%%%%%%%%%

As we explained, we fix the multiplier $\nu$ large (but still much smaller than unity, $\nu = 10^{-5} \ll 1$) to obtain physically realistic PRCs.  Here, to gain insights into how the shapes of the optimal PRCs are determined, we set $\nu=0$ and ignore the 4th-order derivatives in the Euler-Lagrange equations, which does not affect the solutions qualitatively.  With this approximation, the dependence of the optimal solution on the constraint $B$ or on the Lagrange multiplier $\mu$ can be clarified by a simple phase-plane analysis.

\subsubsection{Excitatory impulses}

We set $\nu=0$ to approximate Eq.~(\ref{eq22}) as
\begin{equation}
G''(\theta) =  -\frac{2\mu}{\lambda} G(\theta) \left ( 1 + G'(\theta) \right )^2
\end{equation}
and rewrite this equation as
\begin{equation}
\left \{
\begin{aligned}
G'(\theta) =& H(\theta),\\
H'(\theta) =& -\frac{2\mu}{\lambda} G(\theta) \left ( 1 + H(\theta) \right )^2.\label{eqA4}
\end{aligned}
\right .
\end{equation}
We examine the orbit of this two-dimensional dynamical system as a function of $\theta \in [0,1)$ on the $G-H$ plane with a periodic boundary condition $G(0.5) = G(-0.5)$ and $H(0.5) = H(-0.5)$.

\begin{figure}[tbhp]
  \begin{center}
\includegraphics[width=0.6\hsize]{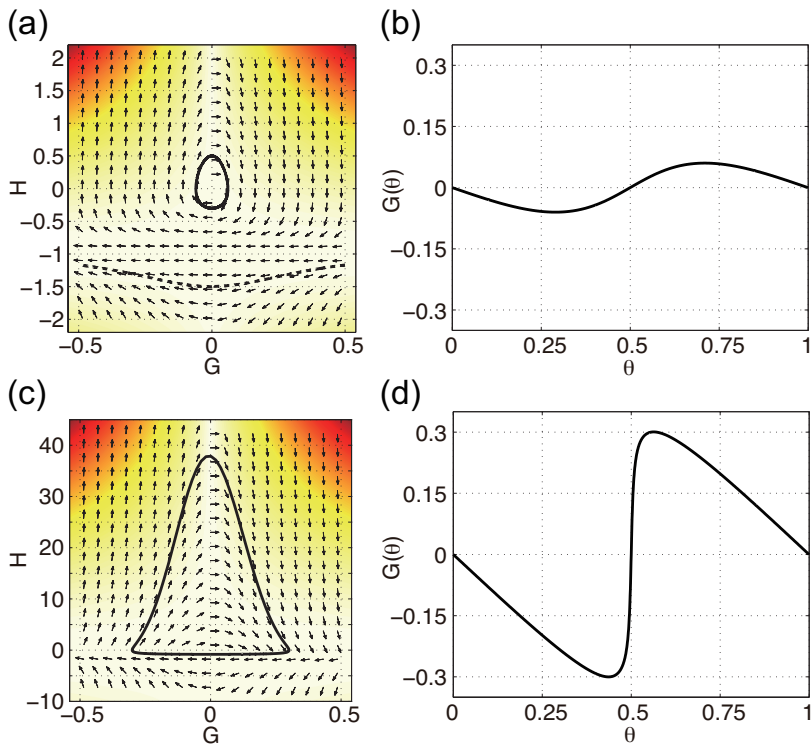}
   \caption{
   (Color online) Vector fields and corresponding PRCs for the case of excitatory impulses.
   Contour color represents magnitude of each vector, where darker color corresponds to higher magnitude.
	(a) Vector field at $\mu = 10$. Solid loop is an orbit with period is $1$.  Broken curve is an orbit starting from $H<-1$.
   (b) Optimal PRCs for $\mu=10$.
   (c) Vector field at $\mu = 15$.  Solid loop plots an orbit whose period is $1$, and (d) the corresponding PRC.
%%   \KILL{(e) Vector field at $\mu = -2.5$.  Solid curve plots an orbit with period $1$, and (f) the corresponding PRC.}
}
   \label{fig05}
  \end{center}
\end{figure}

Let us assume $\mu > 0$ first.  Figure~\ref{fig05}(a) shows an example of the vector field at $\mu = 10$.  The horizontal line $H=-1$ is a separatrix corresponding to the sawtooth solution $G'(\theta)=-1$.
All orbits starting from $H>-1$ are closed, implying the existence of a conserved quantity.  Applying Noether's theorem~\cite{Goldstein} to the Lagrangian in Eq.~(\ref{eq22}), we find that the quantity
\begin{align}
C_{1} = \lambda \left \{  \frac{G'(\theta)}{1+G'(\theta)} - \ln |1+G'(\theta)| \right \} - \mu G(\theta)^{2}
\end{align}
is actually conserved along the flow generated by Eq.~(\ref{eqA4}), reflecting the translational symmetry of the Lagrangian with respect to phase, namely, that the Lagrangian does not depend on $\theta$ explicitly.

A solution possessing period $1$ is chosen from this family of closed orbits by the shooting method.  The solid loop in Fig.~\ref{fig05}(a) shows such a periodic solution, and the solid curve in Fig.~\ref{fig05}(b) is the corresponding optimal PRC.
No orbit starting from $H<-1$ can form a closed loop, because the vector field points to the upper-right and lower right in the third and fourth quadrant, respectively.
Thus, in this region, the orbits have to jump from $G(\theta)=-0.5$ to $0.5$ as shown by a broken curve in Fig.~\ref{fig05}(a).  However, the periods of such orbits are always less than $1$ and therefore solutions with $G'(\theta) < -1$ do not exist.

It can be seen from Eq.~(\ref{eqA4}) that the Lagrangian multiplier $\mu$ determines the time scale of the dynamics in the vertical $H$ direction.  As $\mu$ increases, the vertical dynamics becomes faster, so that the orbit is more strongly attracted to the separatrix $H=-1$ and tends to move along it, as shown in Fig.~\ref{fig05}(c).  Correspondingly, the optimal PRC approaches the sawtooth as shown in Fig.~\ref{fig05}(d).
Note that the separatrix $G'=H=-1$ persists even if $\nu > 0$.  This can be confirmed by taking the limit $G'\rightarrow -1$ in Eq.~(\ref{eq22}), which gives $G''\rightarrow 0$.  Thus, the sawtooth limit also persists in the original system.

When $\mu<0$, we obtained the optimal PRCs for desynchronization as summarized in Appendix D.

%% fig 5 %%

%%%%%%%%%%%%%%%%%%%%%%%%%%%%%%%%%%%%%%%%%%%%%%%%%%%%%%%%%%%%%%%%%%

\subsubsection{Excitatory and inhibitory impulses}

\begin{figure}[tbhp]
  \begin{center}
\includegraphics[width=0.6\hsize]{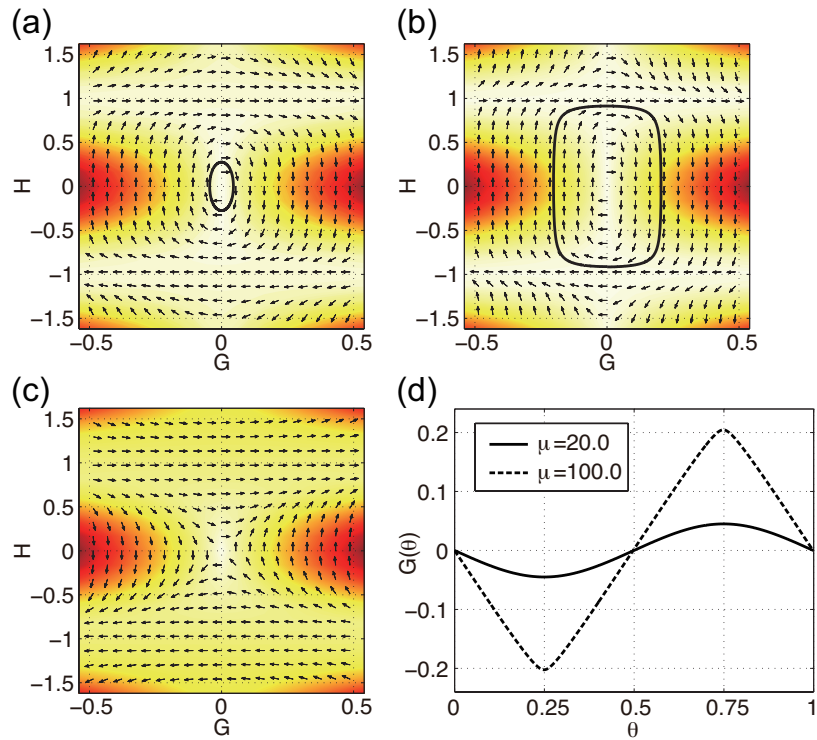}
   \caption{
   (Color online) Vector fields and corresponding PRCs for the case of both excitatory and inhibitory impulses.
   (a) Vector field at $\mu = 20$. Solid loop plots an orbit whose period is $1$.
   (b) Vector field at $\mu = 100$.  Solid loop plots an orbit whose period is $1$.
   (c) Vector field at $\mu = -2.5$.
   (d) Optimal PRCs for two values of $\mu$. Solid curve is the result for $\mu=20$.  Broken curve is for $\mu=100$.}
   \label{fig06}
  \end{center}
\end{figure}

The same analysis can be applied to the case with both excitatory and inhibitory impulses.  As shown in Fig.~\ref{fig06}, when $\mu > 0$, horizontal lines $H = \pm1$ are the separatrices.  Orbits starting from $|H|<1$ always form closed loops, while those starting from $|H|>1$ cannot form a period-$1$ solution.  The conserved quantity in this case is given by
\begin{equation}
C_{2} =  -\lambda \left \{ \frac{G'(\theta)^2}{1-G'(\theta)^2} + \frac{1}{2} \ln \left | 1-G'(\theta)^2 \right | \right \} -\mu G(\theta)^2.
\end{equation}
Increasing the multiplier $\mu$, the optimal solution gradually expands and changes its shape from a circle to a rectangle.  The corresponding PRC deviates from a sinusoid and approaches a double sawtooth.
As before, the separatrices persist even if $\nu > 0$.  No orbit with period $1$ was found when $\mu<0$. 

%% fig 6 %%

%%%%%%%%%%%%%%%%%%%%%%%%%%%%%%%%%%%%%%%%%%%%%%%%%%%%%%%%%%%%%%%%%%

\subsection{Optimal PRCs for stochastic desynchronization}

\begin{figure}[tbhp]
  \begin{center}
\includegraphics[width=0.6\hsize]{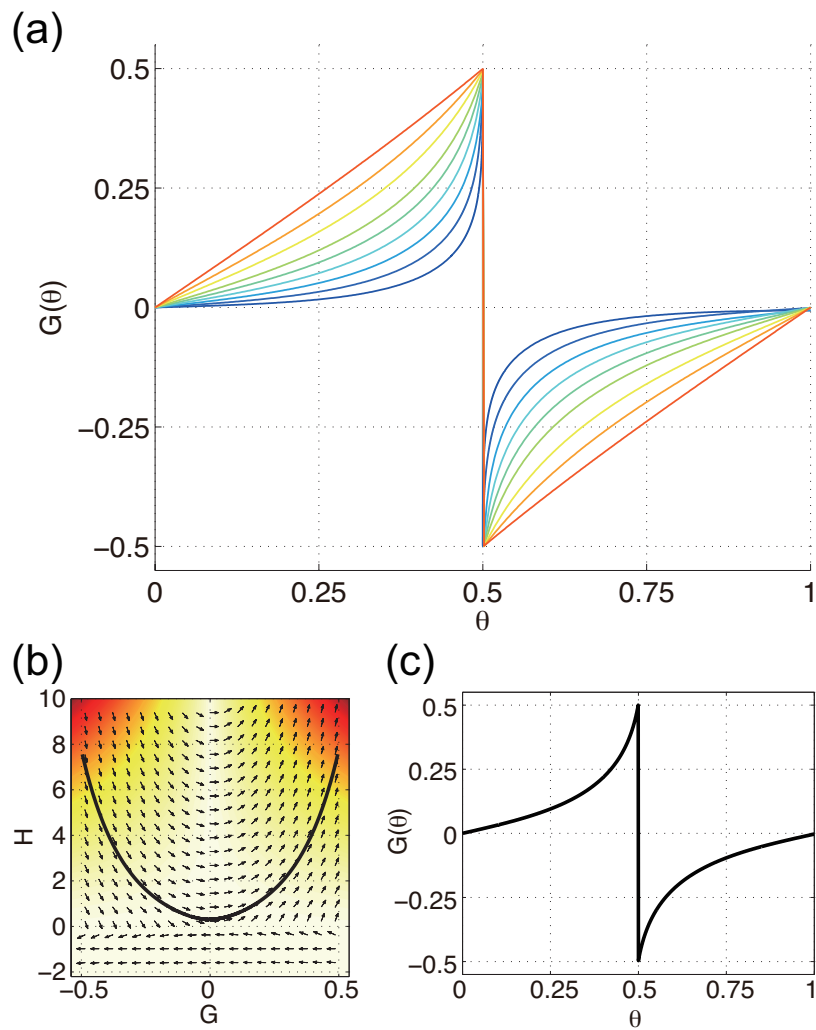}
   \caption{
   (Color online) (a) Optimal PRCs for stochastic desynchronization by excitatory impulses.
   (b) Vector field at $\mu = -2.5$.  Solid curve plots an orbit with period $1$, and (c) the corresponding PRC.
   }
   \label{fig07}
  \end{center}
\end{figure}

The Euler-Lagrange equation gives the solutions that yield the extremum of the action, namely, minimum and maximum, of the Lyapunov exponent $\Lambda$ under the given constraints.
In the case of excitatory impulses, we can vary the Lagrange multiplier $\mu$ controlling the squared amplitude of the PRC in the negative range, $\mu<0$,  while keeping the other Lagrange multiplier $\nu$ the same as in the main text, $\nu = 10^{-5}$, to obtain the PRC that maximizes the Lyapunov exponent.
The corresponding Lyapunov exponent is positive, indicating that the PRC is optimal for stochastic desynchronization~\cite{Arai}.
As shown in Fig.~\ref{fig07}(a), this optimal PRC has a sharp cusp at $\theta=0$ when $\mu$ is sufficiently negative, and gradually approaches a sawtooth as $\mu$ increases.

Examples of the optimal PRC and the corresponding phase-plane orbit are plotted in Figs.~\ref{fig07}(b) and~\ref{fig07}(c).
It is interesting to note that the PRC plotted in Fig.~\ref{fig05}(b) or (d) is ``type 1'' while the PRC in Fig.~\ref{fig07}(a) is ``type 0'' in Winfree's classification~\cite{Winfree1,Winfree2}; the type 1 PRC is continuous and is observed for moderate perturbation intensity, whereas the type 0 PRC is discontinuous and is observed when an oscillator is strongly perturbed~\cite{Czeisler}.
Thus, under the present criteria, the optimal PRC for stochastic synchronization is type 1 and that for desynchronization is type 0.

%% fig 7 %%

%%%%%%%%%%%%%%%%%%%%%%%%%%%%%%%%%%%%%%%%%%%%%%%%%%%%%%%%%%%%%%%%%%


\begin{thebibliography}{99}

%1-5
\bibitem{Mainen} Z. F. Mainen and T. J. Sejnowski, Science {\bf 268}, 1503 (1995).
 
\bibitem{Galan1} R. F Gal\'an, N. Fourcaud-Trocm\'e, G. B. Ermentrout, and N. N. Urban, J. Neurosci. {\bf 26}, 3646 (2006); R. F. Gal\'an, G. B. Ermentrout and N. N. Urban, Sensors and Actuators B: Chemical. {\bf 116}(1-2), 168 (2006); 
R. F. Gal\'an, G. B. Ermentrout and N. N. Urban, J. Neurophysiol. {\bf 99}, 277 (2008); G. B. Ermentrout, R. F. Gal\'an, and N. N. Urban, Trends in Neurosciences {\bf 31}, 428 (2008).

\bibitem{Uchida} A. Uchida, R. McAllister, and R. Roy, Phys. Rev. Lett. {\bf 93}, 244102 (2004).

\bibitem{Yoshida} K. Yoshida, K. Sato, and A. Sugamata, J. Sound and Vibration {\bf 290}, 34 (2006).

\bibitem{Ranta} E. Ranta, V. Kaitala and E. Helle,
Oikos \textbf{78}, 136 (1997).

%6-10
\bibitem{Teramae} J. Teramae and D. Tanaka,
Phys. Rev. Lett. {\bf 93}, 204103 (2004).

\bibitem{Goldobin} D. S. Goldobin and A. S. Pikovsky,
Physica A {\bf 351}(1), 126 (2005).

\bibitem{Goldobin2} D. S. Goldobin and A. S. Pikovsky, Phys. Rev. E {\bf 71}, 045201(R) (2005).

\bibitem{Nagai} K. Nagai, H. Nakao, and Y. Tsubo,
Phys. Rev. E {\bf 71}, 036217 (2005).

\bibitem{Nakao} H. Nakao, K. Arai, K. Nagai, Y. Tsubo, and Y. Kuramoto,
Phys. Rev. E {\bf 72}, 026220 (2005).

%11-15
\bibitem{Arai} K. Arai and H. Nakao, Phys. Rev. E {\bf 78}, 066220 (2008).

\bibitem{Hata} S. Hata, T. Shimokawa, K. Arai, and H. Nakao,
Phys. Rev. E {\bf 82}, 036206 (2010).

\bibitem{Pikovsky} A. Pikovsky, M. Rosenblum, and J. Kurths,
{\it Synchronization} (Cambridge University Press, England, 2001).

\bibitem{Toral} R. Toral, C. Mirasso, E. Hernandez-Garcia, and O. Piro,
Chaos {\bf 11}, 665 (2001).

\bibitem{Zhou} C. Zhou and J. Kurths,
Phys. Rev. Lett. {\bf 88}, 230602 (2002).

%16-20
\bibitem{Nakao2} H. Nakao, K. Arai, and Y. Kawamura,
Phys. Rev. Lett. {\bf 98}, 184101 (2007).

\bibitem{Marella} S. Marella and G. B. Ermentrout,
Phys. Rev. E {\bf 77}, 041918 (2008).

\bibitem{Winfree0} A. T. Winfree, J. Theor. Biol. {\bf 16}, 15 (1967).

\bibitem{Winfree1} A. T. Winfree, {\it The Geometry of Biological Time} (Springer, New York, 2001).

\bibitem{Kuramoto} Y. Kuramoto, {\it Chemical Oscillation, Waves, and Turbulence} (Springer-Verlag, Tokyo, 1984) (republished by Dover, New York, 2003).

%21-25
\bibitem{Holmes} E. Brown, J. Moehlis, and P. Holmes,
Neural Computation {\bf 16}(4), 673 (2004).

\bibitem{Goldbin3} D. S. Goldobin and A. Pikovsky,
Phys. Rev. E {\bf 73}, 061906 (2006).

\bibitem{Khalsa} S. B. S, Khalsa, M. E. Jewett, C. Cajochen, and C. Czeisler, J. Physiol. {\bf 549}, 945 (2003).

\bibitem{Galan2} R. F. Gal\'an, G. B. Ermentrout and N. N. Urban,
Phys. Rev. Lett. {\bf 94}, 158101 (2005).

\bibitem{Tateno} T. Tateno and H. P. C. Robinson, Biophysical Journal {\bf 92}, 683 (2007).

%26-30
\bibitem{Robinson} N. W. Gouwens, H. Zeberg, K. Tsumoto, T. Tateno, K. Aihara, and H. P. C. Robinson, PLoS Comput. Biol. {\bf 6}(9), e1000951 (2010); H. P. C. Robinson, private communication.

\bibitem{Nesse} W. H. Nesse and G. A. Clark,
Biol. Cybern. {\bf 102}, 389 (2010). 

\bibitem{Abouzeid} A. Abouzeid and G. B. Ermentrout,
Phys. Rev. E {\bf 80}, 011911 (2009).

\bibitem{Numerical} W. H. Press, B. P. Flannery, S. A. Teukolsky, and W. T. Vetterling, {\it Numerical Recipes 3rd Edition: The Art of Scientific Computing} (Cambridge University Press, 2007).

\bibitem{Goldstein} H. Goldstein, C. P. Poole, and J. L. Safko, {\it Classical Mechanics} (Addison-Wesley, 2001).

 %31-
\bibitem{Hanson} F. B. Hanson, {\it Applied Stochastic Processes and Control for Jump-Diffusions: Modeling, Analysis, and Computation} (SIAM, 2006).

\bibitem{Lefort} S. Lefort, C. Tomm, J.-C. Floyd Sarria and C. C. H. Petersen, Neuron {\bf 61}, 301-316 (2009).

\bibitem{Winfree2} L. Glass and A. T. Winfree, American Journal of Physiology {\bf 246}, R251-R258 (1984).

\bibitem{Czeisler} C. A. Czeisler, R. E. Kronauer, J. S. Allan, J. F. Duffy, E. N. Brown, and J. M. Ronda, Science {\bf 244}, 1328 (1989).

\end{thebibliography}
\end{document}